\documentclass[11pt]{article}
\usepackage{times}
\usepackage{sidecap}
\usepackage{geometry}
\usepackage{floatrow}
\usepackage{float}
\usepackage{hyperref}
\usepackage{graphics,graphicx}
\usepackage{subcaption}
\usepackage{epsf,wrapfig}
\usepackage{upgreek}
\usepackage{natbib,bm}

\usepackage{paralist}
\let\olditem\item
\renewenvironment{thebibliography}[1]{%
  \section*{\refname}
  \let\par\relax
  \renewcommand{\item}[1][]{\olditem[\textbullet]}%
  \inparaenum}{\endinparaenum}

\newcounter{aff}

\bibpunct{(}{)}{;}{a}{}{,}

\DeclareGraphicsExtensions{.jpg,.pdf,.png,.eps}

\usepackage{caption}
\usepackage{xcolor}

\usepackage{tcolorbox}

\usepackage{soul}

\definecolor{darkcyan}{rgb}{0.0, 0.0, 0.7}

\definecolor{boxcolor}{RGB}{193, 229, 255}
\definecolor{figboxcolor}{RGB}{193, 229, 255}
\definecolor{boxcolorgray}{RGB}{227, 227, 227}
\definecolor{darkcyan}{rgb}{0.0, 0.0, 0.7}

\sethlcolor{boxcolorgray}

\def\gtrsim{\lower 2pt \hbox{$\, \buildrel {\scriptstyle >}\over
{\scriptstyle \sim}\,$}}
\def\lesssim{\lower 2pt \hbox{$\, \buildrel {\scriptstyle <}\over
{\scriptstyle \sim}\,$}}

\def\approxlt{\lower.2em\hbox{$\buildrel < \over \sim$}}
\def\approxgt{\lower.2em\hbox{$\buildrel > \over \sim$}}

\def\be{\begin{equation}}
\def\ee{\end{equation}}
\def\bea{\begin{eqnarray*}}
\def\eea{\end{eqnarray*}}

\def\jref@jnl#1{{\rm#1}}
\def\aj{\jref@jnl{AJ}}
\def\araa{\jref@jnl{ARA\&A}}
\def\apj{\jref@jnl{ApJ}}
\def\apjl{\jref@jnl{ApJ}}
\def\apjs{\jref@jnl{ApJS}}
\def\ao{\jref@jnl{Appl.~Opt.}}
\def\apss{\jref@jnl{Ap\&SS}}
\def\aap{\jref@jnl{A\&A}}
\def\aapr{\jref@jnl{A\&A~Rev.}}
\def\aaps{\jref@jnl{A\&AS}}
\def\azh{\jref@jnl{AZh}}
\def\baas{\jref@jnl{BAAS}}
\def\jrasc{\jref@jnl{JRASC}}
\def\memras{\jref@jnl{MmRAS}}
\def\mnras{\jref@jnl{MNRAS}}
\def\pra{\jref@jnl{Phys.~Rev.~A}}
\def\prb{\jref@jnl{Phys.~Rev.~B}}
\def\prc{\jref@jnl{Phys.~Rev.~C}}
\def\prd{\jref@jnl{Phys.~Rev.~D}}
\def\pre{\jref@jnl{Phys.~Rev.~E}}
\def\prl{\jref@jnl{Phys.~Rev.~Lett.}}
\def\pasp{\jref@jnl{PASP}}
\def\pasj{\jref@jnl{PASJ}}
\def\qjras{\jref@jnl{QJRAS}}
\def\skytel{\jref@jnl{S\&T}}
\def\solphys{\jref@jnl{Sol.~Phys.}}
\def\sovast{\jref@jnl{Soviet~Ast.}}
\def\ssr{\jref@jnl{Space~Sci.~Rev.}}
\def\zap{\jref@jnl{ZAp}}
\def\nat{\jref@jnl{Nature}}
\def\iaucirc{\jref@jnl{IAU~Circ.}}
\def\aplett{\jref@jnl{Astrophys.~Lett.}}
\def\apspr{\jref@jnl{Astrophys.~Space~Phys.~Res.}}
\def\bain{\jref@jnl{Bull.~Astron.~Inst.~Netherlands}}
\def\fcp{\jref@jnl{Fund.~Cosmic~Phys.}}
\def\gca{\jref@jnl{Geochim.~Cosmochim.~Acta}}
\def\grl{\jref@jnl{Geophys.~Res.~Lett.}}
\def\jcp{\jref@jnl{J.~Chem.~Phys.}}
\def\jgr{\jref@jnl{J.~Geophys.~Res.}}
\def\jqsrt{\jref@jnl{J.~Quant.~Spec.~Radiat.~Transf.}}
\def\memsai{\jref@jnl{Mem.~Soc.~Astron.~Italiana}}
\def\nphysa{\jref@jnl{Nucl.~Phys.~A}}
\def\physrep{\jref@jnl{Phys.~Rep.}}
\def\physscr{\jref@jnl{Phys.~Scr}}
\def\planss{\jref@jnl{Planet.~Space~Sci.}}
\def\procspie{\jref@jnl{Proc.~SPIE}}

\def\sun{\hbox{$\odot$}}

\def\ion#1#2{#1$\;${\small\rm\@Roman{#2}}\relax}

\newcount\lecurrentfam
\def\LaTeX{\lecurrentfam=\the\fam \leavevmode L\raise.42ex
\hbox{$\fam\lecurrentfam\scriptstyle\kern-.3em A$}\kern-.15em\TeX}
\def\sizrpt{
(\fontname\the\font): em=\the\fontdimen6\font, ex=\the\fontdimen5\font
\typeout{
(\fontname\the\font): em=\the\fontdimen6\font, ex=\the\fontdimen5\font
}}
\definecolor{dark}{rgb}{ 0.34, 0.09, 0.49}

\usepackage{fancyhdr,url}
\usepackage{transparent}
\usepackage{tablefootnote}
\usepackage{mathpazo}

\begin{document}

\noindent
\begin{raggedright}
{\Huge\bfseries What fuels the growth of stars and black holes in galaxies through cosmic time?
\\[0.5em]
\Large The physics and chemistry of the interstellar medium in distant galaxies
}\end{raggedright}\\

\noindent
Leindert A. Boogaard$^1$,
Manuel Aravena$^{2}$,
Matus Rybak$^{3}$,
Gerg\H{o} Popping$^4$,
Pierre Cox$^5$,
Irene Shivaei$^6$,
Jorge Gonz\'alez-L\'opez$^{7}$,
Jacqueline Hodge$^{1}$,
Karin Cescon$^{1}$,
Francoise Combes$^8$,
Chian-Chou Chen$^{22}$,
Roberto Decarli$^9$,
Miroslava Dessauges-Zavadsky$^{10}$,
Tanio Diaz-Santos$^{11}$,
Andreas Faisst$^{12}$,
Mahmoud Hamed$^{6}$,
Mahsa Kohandel$^{9,13}$,
Adarsh Kumar$^{14}$,
Andrea Lapi$^{15}$,
Umberto Maio$^{16}$,
Dieu D. Nguyen$^{17}$,
Andrea Pallottini$^{13,18}$,
Mamta Pandey-Pommier$^{19}$,
Prachi Prajapati$^{20,21}$,
Dominik Riechers$^{20}$,
Lucie Rowland$^4$,
Hannah Stacey$^4$,
and Livia Vallini$^9$

\vspace{1em}
\emph{Affiliations can be found after the references.}
\vspace{1em}

\begin{tcolorbox}[boxsep=3pt,left=0pt,right=0pt,top=1pt,bottom=1pt,colframe=boxcolor,colback=boxcolor,halign=justify]
  \textbf{Abstract:} The interplay between the baryonic matter in
  galaxies (gas, dust, stars) has a fundamental role in our
  understanding of galaxy formation and evolution through cosmic
  time. The (sub)mm regime provides a unique window to measure the
  physical conditions and chemistry in the cold and dusty gas
  reservoirs that fuel the formation of new stars and the growth of
  massive black holes in distant galaxies.  Large sub(mm)
  interferometers have made a revolutionary impact mapping the cold
  gas and dust in galaxies beyond the local universe.  However,
  significant progress is hampered by limited spectral-line
  sensitivity. After more than a decade of surveys, molecular gas
  (traced by $^{12}$CO) has been measured only in a few tens of
  typical, massive main-sequence star-forming galaxies at $z\geq1$,
  while access to fainter gas-tracing species has been limited to only
  the very brightest galaxies.  Thereby it restricts access to the
  detailed conditions in the cold gas clouds in typical galaxies in
  the distant universe.  A revolutionary, flagship sub-mm
  interferometer with a massive increase in instantaneous spectral
  line sensitivity and broad spectral grasp (ALMA2040) fills this gap
  by enabling rich, multi-transition, multi-species and
  multi-isotopologue studies of the atomic and molecular interstellar
  medium for thousands of typical galaxies across galaxy populations
  and redshift.  This will unveil how the ISM evolves across the
  galaxy population for statistical samples in different redshift
  bins, detailed studies of the physical conditions and chemistry in
  the multi-phase interstellar medium of typical galaxies, and how the
  cold ISM regulates and responds to star- and black-hole formation,
  providing a tremendous leap forward in the study of galaxy assembly
  over cosmic time.
\end{tcolorbox}

\newpage{}

\section*{Scientific content and open questions}

\subsection*{Galaxy evolution and the cosmic cycle of baryons}
\begin{figure}[t]
  \centering
  \includegraphics[width=\textwidth]{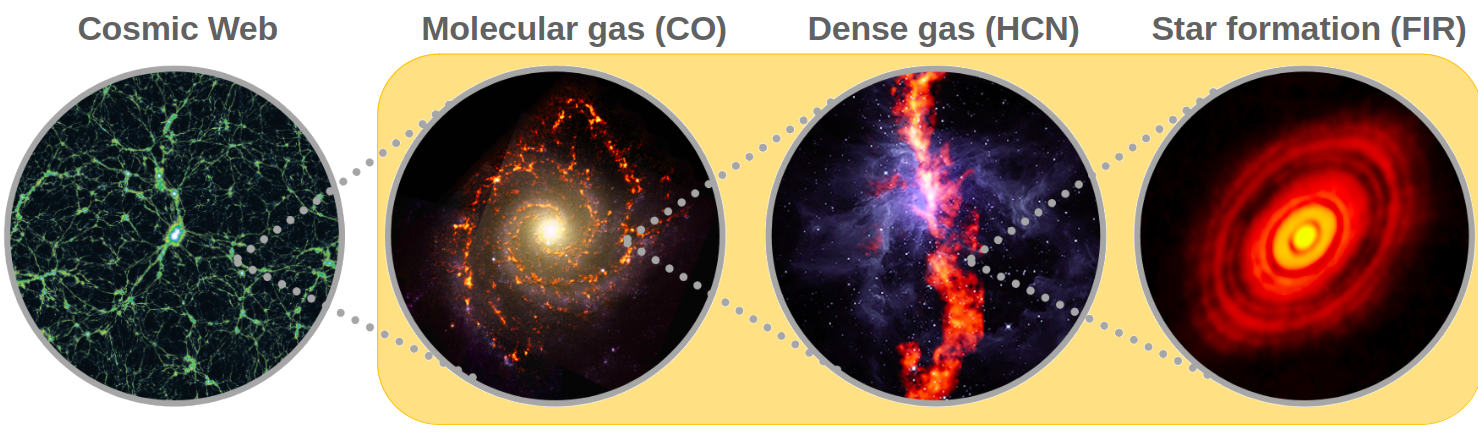}
  \caption{A schematic view of the scales and phases of star formation: from the diffuse gas in the Cosmic Web (traced by HI and nebular lines) to molecular gas (traced by CO), denser gas phases (traced by HCN, HCO$^{+}$, etc.), to protostars. The physical and chemical conditions in the molecular gas clouds---traced by fainter sub-mm emission lines beyond CO---remain largely unexplored beyond the local Universe, due to the limited sensitivity of current mm-wavelength facilities.
    \label{fig:ism-overview}
  }
\end{figure}

The interplay between the baryonic matter in galaxies (gas, dust,
stars) and the formation of stars and growth of supermassive
black-holes plays a fundamental role in our understanding of galaxy
formation and evolution.  Typical galaxies in the early Universe are
observed to be dramatically different from their local counterparts,
with higher star formation densities, lower metal abundances and
larger gas fractions, in environments that strongly evolve over cosmic
time (Madau \& Dickinson 2014; Tacconi et al.\ 2020; Maiolino \&
Mannucci 2019; F\"orster Schreiber \& Wuyts 2020).  But how these
changes impact the physics and chemistry in the cold interstellar
medium (ISM) of galaxies, and the subsequent stellar and black-hole
growth over cosmic time, is still poorly understood.  Key to answering
these questions are both mapping how the gas content of galaxies
varies across the galaxy population, and measuring the physical and
chemical properties inside the cold gas and dust clouds of typical
main-sequence star-forming galaxies at $z>1$, that are uniquely
accessible at (sub)mm wavelengths (~\autoref{fig:ism-overview}).

\subsection*{How do cold gas and dust reservoirs fuel galaxy formation
  through cosmic time?}
\begin{figure}[t]
\centering
\includegraphics[height=7cm]{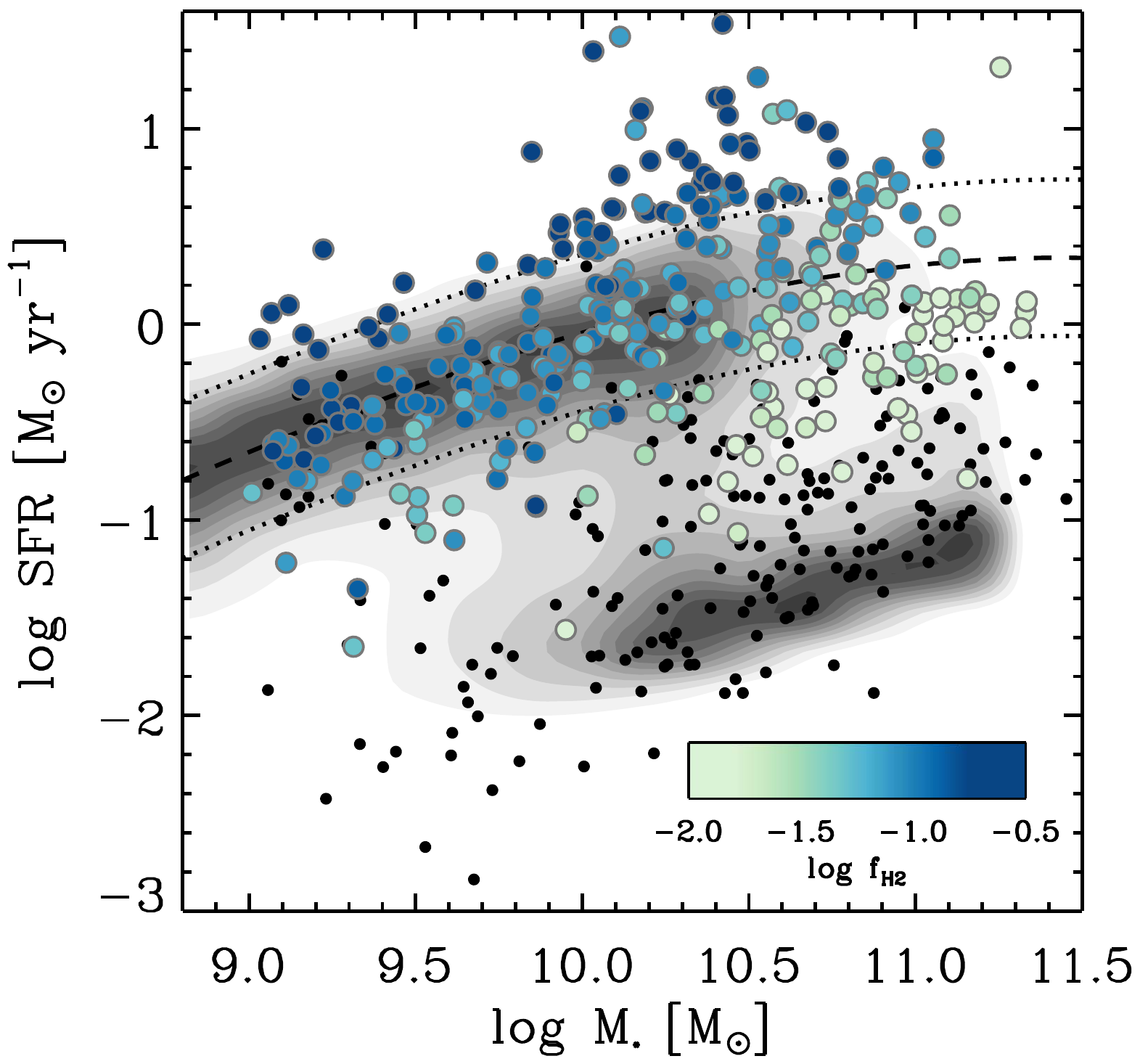}
\includegraphics[height=7cm]{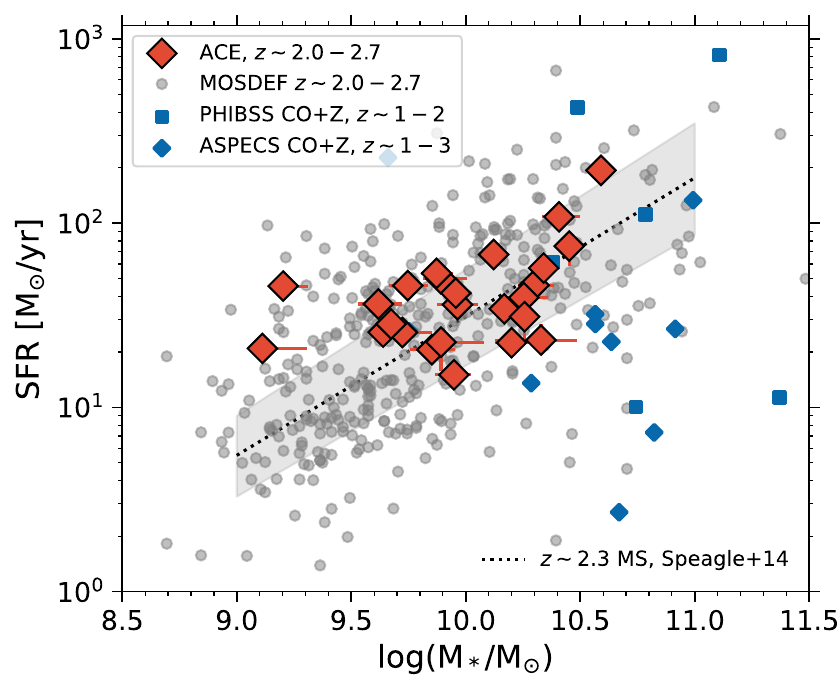}
\caption{\textbf{Left:} In the local universe, molecular gas fractions
  have been measured in the $\rm SFR-M_*$ plane for hundreds of
  galaxies with rest-optical spectroscopy from the SDSS across the
  galaxy main sequence ($z=0$; Saintonge et al.\
  2017). \textbf{Right:} In contrast, surveys at $z>1$ are still
  plagued by small sample sizes, with three large programs spanning
  15\,years (red \& blue points) having together yielded molecular gas
  measurements in $<50$ typical main-sequence galaxies with known
  metallicities in the broad redshift interval $z=1$--3 (Shivaei et
  al.\ 2026) and only a handful at higher redshifts (e.g. Cescon et
  al.\ 2026).  \textbf{ALMA2040 will be able to measure the molecular
    gas content of thousands of galaxies at \textit{z}=1--6 (and
    beyond) in discrete redshift intervals, and map how the fuel for
    stellar and black hole formation evolves across the population.}}
  \label{fig:galaxies}
\end{figure}

Cold molecular gas clouds consist predominantly of H$_2$ that
practically does not radiate at low temperatures (e.g., Bolatto et
al.\ 2013).  Instead, the workhorse tracers of the cold molecular gas,
that fuels the star formation, are the bright rotational transitions
of carbon monoxide ($^{12}$CO).  Surveys in the local universe have
revealed how the molecular gas reservoirs of galaxies evolve across
the population (e.g., Saintonge et al.\ 2017).  These have revealed
that the molecular gas content varies dramatically with mass, star
formation rate and metallicity.  However, while after more than three
decades of observations the total number of $^{12}$CO detections at
$z>1$ is now a few hundred, these largely consist of the most
IR-luminous sources including (high-redshift) quasars and
submillimeter galaxies (Carilli et al.\ 2013, Tacconi et al.\ 2020).
In contrast, the number of more `typical' main-sequence star-forming
galaxies detected in $^{12}$CO at $z>1$ is less than a hundred, with
only a handful beyond $z>3$ (Aravena et al.\ 2020, 2024, Cescon et
al.\ 2026, Shivaei et al.\ 2026).  Furthermore, even these samples of
more `typical' galaxies are largely limited to the massive and
gas/dust rich end of the galaxy population
($M_* \geq 10^{11}$\,M$_{\odot}$), while the number of lower-mass
galaxies is even fewer (see \autoref{fig:galaxies}).  This is not
surprising, as while lower-mass galaxies are relatively more gas rich,
the gas content of galaxies at fixed redshift to first order scales
with their stellar mass and star formation rate (see Tacconi et al.\
2020); hence, probing lower mass galaxies requires a corresponding
increase in spectral line sensitivity in the observations.  Clearly,
the samples of galaxies assembled to date still give a biased view of
the overall galaxy population at $z\geq1$ and demonstrate the need for
representative studies probing larger samples of more normal, lower
mass and star formation rate galaxies.  In addition, the small numbers
stand in stark contrast with the thousands of galaxies that will be
mapped by the JWST, ELT and large ground- and space-based surveys by
the 2040's, and will provide unprecedented constraints on the stellar
populations and ionized gas in distant galaxies.

Moreover, measuring the physical conditions and chemistry in the cold
gas and dust, which are expected to be dramatically different in early
galaxies, critically requires mapping the emission from molecules
beyond $^{12}$CO.  These include more complex and denser gas-tracing
molecules (HCN, HCO$^{+}$, etc.), the optically thin isotopologues of
$^{12}$CO ($^{13}$CO, C$^{18}$O) and other molecules, as well as
(neutral, atomic) fine-structure lines ([CI], [OI]), in both their
ground- and excited states, that are only detectable at (sub)mm
wavelengths.  However, as these lines are at least $\geq$5--10$\times$
fainter than $^{12}$\,CO, detecting them is currently practically
impossible for typical galaxies even at cosmic noon ($z=1$--3, for
which even the bright $^{12}$CO and [CI] lines have only been scarcely
mapped; e.g., Boogaard et al.\ 2020, Valentino et al.\ 2020) and this
will remain so even after the upcoming ALMA Wide Band Sensitivity
Upgrade, except for the most extremely luminous and/or strongly lensed
galaxies (e.g., Yang et al.\ 2023).  This imposes a fundamental
limitation on our understanding of the physics of stellar and
black-hole growth, and how it is impacted by cosmic environment and
galaxy evolutionary stage, from the early universe into the epoch of
reionisation ($z\geq6$) to the peak of cosmic star formation
($z$=1--3).

\subsection*{The need for a next-generation (sub)mm interferometer}
The lack of sensitivity needed to access beyond the brightest emission
lines and to obtain statistically large samples of galaxies has
fundamentally limited our understanding of the physics of stellar and
black-hole growth, and how it is impacted by galaxy environment and
evolution, from the early universe ($z\geq6$) to the peak of cosmic
star formation ($z$=1--3).  A transformational next step requires an
order-of-magnitude increase in spectral line sensitivity on galaxy
scales.  This can only be achieved with a next-generation ALMA-like
interferometer (ALMA2040).

\section*{ALMA2040: Key science cases}

\subsection*{Mapping the molecular gas content for large galaxy
  samples through cosmic time}

A major next step in constraining galaxy formation and evolution
models will come from measuring the cold gas reservoirs---the fuel for
star formation---across the typical galaxy population.  This includes
in particular probing the lower mass and metallicity regime, that
plays a fundamental role in hierarchical galaxy growth, as well as
mapping the impact of galaxy environment and evolutionary stage.

Efficient surveys with ALMA2040 of the bright, workhorse molecular gas
tracers at $z>1$ (incl.\ $^{12}$CO, [CI]) for large samples of
galaxies on integrated scales will be key to mapping the gas content
of galaxies across the galaxy population.  These reveal how the
molecular gas fraction and depletion times vary across main sequence
galaxies for statistical samples beyond the local Universe.  Larger
surveys will be able to assess how the gas content evolves across
environments, in comparison to theoretical model predictions, and
catch rarer and shorter-lived galaxy evolutionary stages.  Deeper
surveys in particular will also open a window on the molecular gas in
truly low-mass, dwarf and metal-poor galaxies at higher redshift.
These observations will address key open questions such as what drives
starbursts in galaxies at different cosmic epochs and, on the opposite
side, what processes drive the shutdown of star formation, sometimes
already very early in their evolution (see associated white paper by
Zanella, Belli, et al.).

\autoref{fig:spectrum} shows simulated spectra of star-forming
galaxies at $z\sim3$ from simulations for a fiducial ALMA2040 like
array.  Specifically, these sensitivity estimates correspond to an
approximately ten times more sensitive array of $\approx$200--300
antennas plus an anticipated increase in the overall sensitivity due
to improvements in uv-coverage, receivers, correlator, backend, etc.,
compared to ALMA.  The spectra reveal the wealth of sub-mm emission
lines, of which the strongest lines of $^{12}$CO can already be
detected even in shallow ($<$1\,hour) exposures.  Based on the number
densities of typical galaxies in the sub-mm around 250\,GHz (e.g.,
Gonzalez-Lopez et al.~2020), and experience from deep exposures with
ALMA, every individual ALMA2040 pointing will yield multiple sources
in the field of view, immediately obtaining broad spectral coverage of
multiple lines. As a result, large numbers of galaxies can be surveyed
highly efficiently.

The locally-calibrated CO-to-H$_{2}$ mass conversion factors, that
underpin most of our understanding of the cold gas content of distant
galaxies, may potentially evolve significantly in different
environments at high redshift (Bolatto et al.\ 2013).  Therefore,
simultaneous access to multiple CO lines and continuum will be
absolutely critical to assess the global gas excitation and thereby
limit the current uncertainties driven by single gas tracers.  Key to
achieving this is a broad spectral grasp and sufficient sensitivity to
capture multiple lines in single, few-hour pointings.  Sufficient
angular resolution to deblend individual galaxies ($\lesssim 1.5''$ at
3\,mm and even longer wavelengths) is crucial to probe down to low
masses where often clustered galaxies are spectrally hard to
disentangle.  These observations can subsequently leverage deeper
studies for fewer galaxies (see next section) that will unlock the
fainter ISM tracers, providing a more comprehensive view of the
physical and chemical state of the cold gas in distant galaxies.

\subsection*{Measuring the ISM chemistry and physics through
  fainter gas tracers beyond
  $^{12}$CO} \label{sec:physchem}

\begin{figure}[h!]
  \centering
  \includegraphics[width=\textwidth]{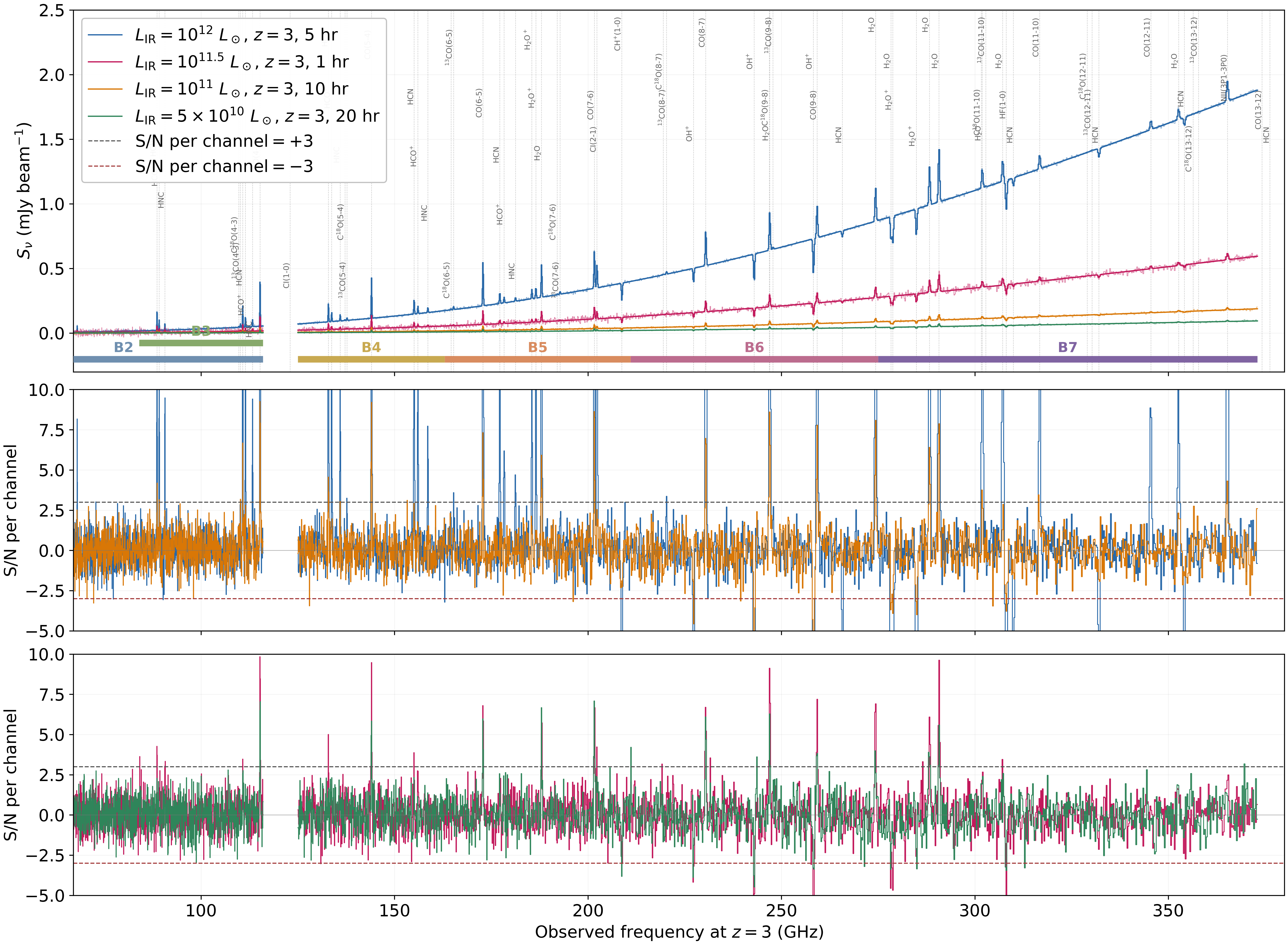}
  \caption{Simulated ALMA2040 spectra for $z=3$ galaxies of different
    total infrared luminosity ($L_{\rm IR}$) roughly corresponding to
    starburst ($10^{12}$\,$L_{\sun}$), a main-sequence
    ($10^{11}$\,L$_{\sun}$), and lower SFR galaxies
    ($10^{10.5}$\,L$_{\sun}$) at $z\sim3$.
    The spectrum is based on an
    observed nearby starburst galaxy template (Rangwala et al.\
    2011; with some fainter lines not originally measured have been added
    based on radiative transfer model predictions/scalings from the
    detected lines) and assumes a line width and spectral resolution
    of 400 and 120\,km\,s$^{-1}$, respectively.  We note that this
    template has relatively strong continuum and in reality line
    ratios and equivalent widths will vary for galaxies of different
    $L_{\rm IR}$.  \textbf{A moderately deep integration
      ($\mathbf{\sim}$5\,hr) with ALMA2040 could simultaneously
      capture all bright lines (CO, [CI], H$_2$O) as well as denser
      gas tracing species (e.g., HCN, HCO$^{+}$, etc.), various
      absorption lines tracing feedback activity (e.g., OH$^+$,
      CH$^+$, HF), and faint isotopologues for galaxies at $z\sim3$;
      capturing a significant fraction of the excitation ladder of
      multiple species as well as the underlying dust continuum in a
      single shot.}  Note the total bandwidth shown in the figure is
    (only) $\sim$300\,GHz while ALMA2040 should extend to both lower
    (ALMA bands 1--2) and higher (ALMA bands 8--10) frequencies not
    shown in the plot.  \textbf{In addition, given the extremely broad
      instantaneous bandwidth foreseen for ALMA2040
      ($\mathbf{\geq}$512\,GHz), the efficiency of these observations
      will be hundreds of times higher than with ALMA,
      which even after the full WSU upgrade will require many
      ($\mathbf{\geq}$25) tunings to scan all of Bands 1--8
      ($\mathbf{\approx}$425\,GHz) and a similar number for Bands 9+10
      ($\mathbf{\approx}$280\,GHz).}
    \label{fig:spectrum}}
  \vspace{\baselineskip} 
\end{figure}

Tracing the physical conditions and chemistry inside the cold gas
reservoirs of typical galaxies through cosmic time will be a crucial
leap in our understanding of how the gas in the ISM is converted into stars and fuels
black-hole formation.  Sensitive multi-frequency spectral line
inventories will unlock the physics and chemistry of cold gas in
galaxies at $z>1$ that are currently inaccessible in all but the
brightest galaxies at high redshift.

More complex molecular species beyond CO (such as HCN, HCO$^+$, etc.)
have higher critical densities and thus trace the denser gas phases
that are more directly connected with star formation.  While in the
local Universe these indeed correlate closely (Gao \& Solomon 2004),
and are used to trace dense gas fraction and star-formation
efficiency, initial studies in the brightest high-redshift systems
suggest these may deviate significantly (e.g., Oteo et al.\ 2017;
Rybak et al.\ 2022, 2026) leaving it unclear if local calibrations
apply to more typical systems at high redshift.  In addition to the
dense gas tracers, transitions from water will be ubiquitously
detected throughout in galaxy spectra (see \autoref{fig:spectrum}).
These observations will also capture various molecules and ions
tracing outflows (incl. OH${^+}$, CH$^{+}$, HF; see associated white
paper by Butler, Spilker, Fujimoto, et al.), thereby providing key
insight into the intense radiation field and outflow driving
star-formation and AGN feedback (e.g., Butler et al.\ 2021; Riechers
et al.\ 2021).

Crucially, ALMA2040 could finally give access to optically thin line
isotopologues, in particular $^{13}$CO and C$^{18}$O, for massive
star-forming galaxies on and above the main sequence.  These will
provide measurements of, and break key degeneracies with, the column
densities and excitation conditions in the large gas reservoirs.  This
will constrain enrichment timescales and, jointly with
chemical-evolution models, provide sensitivity to the IMF.  More
practically, this will be key to constrain how the
$^{12}$CO-to-H$_{2}$ conversion factor varies across galaxies with
different metallicities and through cosmic time.  The latter will
leverage important breakthroughs to improve single-species studies of
the gas content of galaxies as well as the evolution of the cosmic gas
density in the context of the baryon cycle (see associated white paper
by Aravena, Boogaard, et al.).

The simulations from \autoref{fig:spectrum} show that for galaxies
with $L_{\rm IR} = 10^{12}$\,L$_{\odot}$, a few hours of observation ($\sim$5\,hr) with
ALMA2040 will reveal a rich inventory of molecular gas lines.  For
more typical main-sequence galaxies at $z\sim3$ with
$L_{\rm IR} \leq 10^{11}$\,L$_{\odot}$, the dense gas lines and even
isotopologues are detectable, though the latter would require deeper
integrations on the order of 10+\,hours.  Excitingly, the CO and
brightest denser gas tracers can even be detected in a deeper
integration for low SFR galaxies at $z\sim3$ that host only a few
M$_{\rm \odot}$~yr$^{-1}$ of star formation
($L_{\rm IR}\approx10^{10.5}$) similar to the Milky Way.

We note that, in addition to the cold-gas related atomic and molecular
species in focus here, many other bright and faint fine structure
lines from the neutral and ionized atoms of various species (incl.\ C,
N, O, Ne, S, Si, Ar, etc.) will be efficiently detected across the
frequency range.  These are discussed in several companion white
papers (e.g.,\ Diaz-Santos, Vallini, et al.).

What is clear is that the spectral lines associated with the molecular
complexity are over an order-of-magnitude fainter than the workhorse
lines measured by ALMA today.  Detecting them routinely through cosmic
time is therefore the unique territory of a next-generation (sub)mm
ALMA-like interferometer.  With ALMA, even after the WSU upgrade,
spectral line inventories at cosmic noon will remain confined to the
brightest, most massive, and often lensed galaxies.  With ALMA2040's
order-of-magnitude gain in spectral-line sensitivity and broad
bandwidths, the molecular line inventory will become accessible to the
bulk of the galaxy population dominating the cosmic star formation and
black-hole growth.  This will yield crucial insight into the molecular
inventories of galaxies through cosmic time and the origins of molecular
complexity.

\section*{Synergies with current and future facilities and simulations}

\subsection*{Multi-phase and multi-facility with ELT, ngVLA and beyond}

ALMA2040 will operate alongside several other powerful and highly
complementary facilities of the 2040s.  The Extremely Large Telescope
(ELT) will deliver (sub-)kpc-scale rest-frame optical/near-infrared
spectroscopy of the ionized ISM, stellar populations, and dynamics of
galaxies whose cold gas phase will be characterized by ALMA2040 (see,
e.g., the white paper by Kaasinen, Kohandel, Rizzo, Bacchini, et al.).
This will yield a key self-consistent, multi-phase picture of the ISM
linking the ionized, atomic, and (dense) molecular gas phases directly
to the underlying stellar populations and to the histories of
star-formation and black-hole-growth that shaped them, building on the
work from current and anticipated facilities (incl. JWST, Euclid,
Roman, PRIMA; as well as upcoming large spectroscopic surveys).  In
the longer-wavelength radio-regime, several new facilities are on the
horizon. The ngVLA will provide a highly complementary view to
ALMA2040, in particular by tracing the ground state molecular
transitions out to (very) high redshifts.  The SKA and DSA (see
Hallinan et al.\ 2020) will be providing a comprehensive picture of
the neutral atomic gas (HI).  Together with ALMA2040, these facilities
will assemble a complete picture of the cosmic baryon cycle.

\subsection*{Connection to new generation simulations of galaxy formation}
By 2040, we can also expect significant progress in numerical studies
of galaxy evolution.  Modeling the coldest gas phases of galaxies
self-consistently remains a key challenge, which state-of-the-art
simulations are increasingly addressing both in zoom-in simulations of
individual galaxies as well as in a cosmological context (e.g., Maio
et al.\ 2022, Pallottini et al.\ 22, Feldmann et al.\ 2023, Schaye et al.\ 2025).  In
addition, key physical processes related to star formation, stellar winds, as well as black hole growth and AGN feedback, remain unresolved in modern
galaxy evolution simulations, which have significant impact on the
overall properties of the gas and the gas flows.  Other processes are absent or included despite large uncertainties (such as the effects of cosmic rays, dust evolution, and magnetic fields).  Thus, to test the physical recipes that underpin these processes, the models and their more mature descendants by 2040 will rely critically on precise
observational constraints on the gas and dust content across galaxies,
and the conditions and abundances of metals and dust and molecules in
the cold gas through cosmic time.

\section*{ALMA2040: Technical requirements}
Addressing these key scientific questions needs a next-generation ALMA
(sub)mm interferometer (ALMA2040) with a transformative,
order-of-magnitude increase in the instantaneous spectral line
sensitivity.  As the lines associated with molecular complexity are at
least an order-of-magnitude fainter than the current workhorse lines
observed with ALMA, this means a $\geq$10x increase in instantaneous
(spectral line) sensitivity compared to current ALMA (or $\sim$8x
compared to the anticipated ALMA Wide Band Sensitivity Upgrade; ALMA
WSU).

Measuring the physics and chemistry of the interstellar medium relies
fundamentally on multi-transition measurements of different excitation
states which lie across the (sub)mm spectrum, even for a single
molecule (for example, for CO the rotational levels are spaced by
$\approx$115\,GHz/$(1+z)$, that is, across many current ALMA bands
even at higher redshifts; cf. \autoref{fig:spectrum}).  Therefore,
broad instantaneous bandwidths will be transformational, as they will
move the field away from the current ``one-transition-at-a-time
surveys'', to instantaneous multi-line surveys, that will be hundreds
of times more efficient than with current ALMA.  The bandwidth should
span across the entire (sub)mm window, that is, at wavelengths from
9\,mm to 300\,$\upmu$m (ALMA bands 1–-10; 35–-950\,GHz).  The
high-frequency capabilities are exceptional at the ALMA site---and not
provided by any other foreseen (sub)mm facility---and the broad
frequency coverage is critical to consistently access the ground and
excited state of key species in the interstellar medium across
redshifts.

Given the expected high multiplicity of low-mass galaxies due to the
increase in sensitivity, sufficient angular resolution at low
frequencies is critical, and the angular resolution should be
sufficiently high to deblend individual galaxies even at the low
frequency end.  At the same time, the lowest achievable angular
resolution should be low enough such that an unresolved observation
can be obtained of a single massive star-forming galaxy at cosmic
noon.  This translates to a lowest resolution of $\theta\approx 1.5''$
($\approx15$\,kpc at cosmic noon); broadly matching the global
diameter of the gas reservoir in massive, gas-rich star-forming
galaxies cosmic noon.  Access to higher angular resolution down to
cloud scales is key to connect to the gas on sub-kpc scales, which are
described in several companion white papers (e.g., Zanella, Faisst and
Herrera-Camus et al.)

These requirements can be achieved by increasing the collecting area
compared to ALMA-WSU (providing much better image fidelity through the
increased $uv$-coverage) and disruptive developments in receiver
technology.  Innovative developments in receiver and signal processing
technology, such as simultaneous multi-band
capabilities and ultra-wide bandwidths (full-band receivers), will enable fast surveying capabilities. Through dramatic improvements in processing speed, the correlator
is no longer a limiting factor for processing the wide
(multi)bandwidth and high spectral resolution data for a large number
of antennas. No increase in the spectral resolution or
sensitivity-limited polarisation purity beyond the ALMA WSU is
currently required.

\section*{Acknowledgements}
This work was developed as part of the broader ALMA2040 community
effort.  We acknowledge the Lorentz Center and NOVA for supporting the
community workshops and coordination activities that helped initiate
and shape the ALMA2040 White Paper process.  We also thank the
external and internal reviewers for their feedback.
L.A.B. acknowledges support from the Dutch Research Council (NWO)
under grant VI.Veni.242.055
(\url{https://doi.org/10.61686/LAJVP77714}).

\section*{References}
Aravena et al.\ 2024, A\&A, 682, 24 $\bullet$
Boogaard et al.\ 2020, ApJ, 902, 109 $\bullet$
Boogaard et al.\ 2023, ApJ, 945, 111 $\bullet$
Butler et al.\ 2021, ApJ, 919, 9 $\bullet$
Cescon et al.\ 2026, MNRAS, 549, 3$\bullet$
Decarli et al.\ 2019, ApJ, 882, 138 $\bullet$
Feldmann et al.\ 2023, MNRAS, 522, 3831 $\bullet$
F\"orster Schreiber \& Wuyts 2020, ARA\&A, 58, 661 $\bullet$
Gao \& Solomon 2004, ApJ, 606, 271 $\bullet$
Gonzalez-Lopez et al.\ 2020, ApJ, 897, 91 $\bullet$
Hallinan et al.\ 2020, arXiv, 2907.07648 $\bullet$
Maio et al.\ 2022, A\&A, 657, A47 $\bullet$
Madau \& Dickinson 2014, ARA\&A, 52, 415 $\bullet$
Maiolino \& Mannucci 2019, A\&ARv, 27, 3 $\bullet$
Oteo et al.\ 2017, ApJ, 850, 170 $\bullet$
Pallottini et al.\ 2022, $\bullet$
Peroux \& Howk 2020, ARAA, 58, 363 $\bullet$
Rangwala et al.\ 2011, ApJ, 743, 94 $\bullet$
Riechers et al.\ 2021, ApJ, 913, 141 $\bullet$
Rybak et al.\ 2022, A\&A, 667, 70 $\bullet$
Rybak et al.\ 2026, A\&A, 706, 69 $\bullet$
Saintonge et al.\ 2017, ApJS, 233, 22 $\bullet$
Schaye et al.\ 2025, arXiv, 2508.21126 $\bullet$
Shivaei et al.\ 2026,arXiv, 2609.21604  $\bullet$
Tacconi et al.\ 2020, ARAA, 58, 157 $\bullet$
Valentino et al.\ 2020, ApJ, 890, 24 $\bullet$
Walter et al.\ 2020, ApJ, 902, 111 $\bullet$
Yang et al.\ 2023,  A\&A, 689, 95

\section*{Affiliations}
{\footnotesize
\noindent $^1$ Leiden Observatory, Leiden University, NL-2300 RA Leiden, The Netherlands\\
$^2$ Instituto de Estudios Astrofisicos, Universidad Diego Portales, Av. Ejercito 441, Santiago, Chile \\
$^4$ European Southern Observatory (ESO), Germany, Karl-Schwarzschild-Str. 2, D-85748, Garching, Germany\\
$^5$ Institut d'Astrophysique de Paris, France, 98bis bd. Arago, 75014 Paris, France\\
$^6$ Centro de Astrobiolog\'ia (CAB), CSIC-INTA, Ctra. de Ajalvir km 4, Torrej\'on de Ardoz, E-28850, Madrid, Spain\\
$^7$ Instituto de Astrofísica, Pontificia Universidad Católica de Chile, Av. Vicuña Mackenna 4860, 7820436, Santiago
Chile\\
$^8$ Observatoire de Paris, LUX, 77 Av. Denfert-Rochereau, F-75014 PARIS, France\\
$^9$ INAF -- Osservatorio di Astrofisica e Scienza dello Spazio, Via P. Gobetti 93/3, I-40129, Bologna, Italy \\
$^{10}$ Department of Astronomy, University of Geneva, Chemin Pegasi 51, 1290 Versoix, Switzerland \\
$^{11}$ Institute of Astrophysics - FORTH, Greece, Heraklion 70013, Greece\\
$^{12}$ PAC, California Institute of Technology, 1200 E. California Blvd. Pasadena, CA 91125, USA, \\
$^{13}$ Scuola Normale Superiore, Piazza dei Cavalieri 7, 56126 Pisa, Italy\\
$^{14}$ Instituto de Astronomía y Ciencias Planetarias, Universidad de Atacama, Copayapu 485, 1531772, Copiapó, Chile\\
$^{15}$ SISSA, Via Bonomea 265, 34136 Trieste, Italy\\
$^{16}$ INAF – Osservatorio Astronomico di Trieste, Via G. Tiepolo 11, 34143 Trieste, Italy\\
$^{17}$ Department of Astronomy, University of Michigan, 1085 South University Avenue, Ann Arbor, MI 48109, USA\\
$^{18}$ Dipartimento di Fisica “Enrico Fermi”, Universitá di Pisa, Largo Bruno Pontecorvo 3, Pisa I-56127, Italy\\
$^{19}$ Pole Scientific, University Catholic of Lyon- University of Lyon, 10 place des Archives 69002, Lyon, France \\
$^{20}$ Institute for Astrophysics, University of Cologne, Z{\"u}lpicher Straße 77, 50937 Cologne, Germany\\
$^{21}$ Max Planck Institute for Radio Astronomy (MPIfR), Auf dem H{\"u}gel 69, 53121 Bonn, Germany\\
$^{22}$ Academia Sinica Institute of Astronomy and Astrophysics (ASIAA), No. 1, Sec. 4, Roosevelt Road, Taipei 106319, Taiwan \\
}
\end{document}